\newif\ifanon\anonfalse  
\def\year{2021}\relax
\documentclass[letterpaper]{article} 
\usepackage{aaai21}  
\usepackage{times}  
\usepackage{helvet} 
\usepackage{courier}  
\usepackage[hyphens]{url}  
\usepackage{graphicx} 
\usepackage{amsthm}
\usepackage{natbib}  
\usepackage{caption} 
\theoremstyle{definition}
\newtheorem{exmp}{Example}
\theoremstyle{definition}
\newtheorem{rmk}{Remark}
\title{Unification of HDP and LDA Models for Optimal Topic Clustering of Subject Specific Question Banks}

\ifanon
\author{Anonymous}
\else
\author {

        Nikhil Fernandes\textsuperscript{\rm 1},
        Alexandra Gkolia\textsuperscript{\rm 1},
        Nicolas Pizzo\textsuperscript{\rm 1},
        James Davenport\textsuperscript{\rm 1}, \\
        Akshar Nair\textsuperscript{\rm 1}\\
}
\affiliations {
    \textsuperscript{\rm 1}University of Bath \\
    nkdf20@bath.ac.uk, ag2214@bath.ac.uk, np700@bath.ac.uk, masjhd@bath.ac.uk, asn42@bath.ac.uk
}
\fi
\begin{document}

\maketitle

\begin{abstract}
There has been an increasingly popular trend in Universities for curriculum transformation to make teaching more interactive and suitable for online courses.
An increase in the popularity of online courses would result in an increase in the number of course-related queries for academics. This, coupled with the fact that if lectures were delivered in a video on demand format, there would be no fixed time where the majority of students could ask questions. When questions are asked in a lecture there is a negligible chance of having similar questions repeatedly, but asynchronously this is more likely.  
In order to reduce the time spent on answering each individual question, clustering them is an ideal choice. There are different unsupervised models fit for text clustering, of which the Latent Dirichlet Allocation model is the most commonly used. We use the Hierarchical Dirichlet Process to determine an optimal topic number input for our LDA model runs. Due to the probabilistic nature of these topic models, the outputs of them vary for different runs. The general trend we found is that not all the topics were being used for clustering on the first run of the LDA model, which results in a less effective clustering. To tackle probabilistic output, we recursively use the LDA model on the effective topics being used until we obtain an efficiency ratio of 1. Through our experimental results we also establish a reasoning on how Zeno’s paradox is avoided.
\end{abstract}

\section{Introduction}
One of the main problems faced by academics with the increase in online courses is to answer students queries. 
\ifanon \else{}
For example, the University of Bath has introduced five new MSc courses as of last year, which introduces a total of over 200 new Masters students as well. \fi The current approach uses a community-based question-answer format, such as StackOverflow, Quora, StackExchange, Google Groups, Blackboard etc. All these methods would require the academic in charge of the course to go through all the questions, answer them or decide if they are linked to any already answered question. To counter this problem, we propose and investigate clustering of question-answer pairs from various databases\ifanon at our university \else{} from the Department of Computer Science at the University of Bath\fi.  

There has been extensive research done on text clustering algorithms. These mainly include unsupervised machine learning methods such as Latent Dirichlet Allocation (LDA), $k$-means, Support Vector Machines, etc.
There is an ongoing debate about the accuracy of these methods due to them being unsupervised. \citet{Hannahetal09} investigated various models focusing on describing the probabilistic nature of these methods, linking to their dependability. They suggest a method that compares the predictive probability of the second half of the document given the first half. 

LDA \cite{lda2003} is a generative statistical model which seeks to express a given observed data set in terms of a mixture of unobserved 'latent' groups. These latent groups are used to capture the notion of 'topics' or 'clusters', and can highlight some hidden structure within the data set. From this, the division of the data set can be used for classification of any new data objects. 

For clarification, when viewing a data set to be fitted with an LDA model, the following can be interpreted as: 'word' refers to the basic unit of data within the data set, 'document' is a collection of words, and 'corpus' is a collection of documents.

We focus on improving the efficiency of the LDA model by optimising the inputs of the LDA algorithm. Given that the LDA model is probabilistic,  not all topics estimated as input will be used to form clusters of the document. To counter this problem, we propose a recursive method that takes advantage of the Hierarchical Dirichlet Process, and recursively input the effective number of topics used in our model.  

 We primarily work on the data set obtained from \ifanon our University \else the Data Structures course offered at the University of Bath\fi. Section `The Dataset' focuses on pre-processing our data set in order to increase the accuracy of the clustering of the LDA model. Further to this, we investigate various permutations (of the order of questions) of our data set and provide an in-depth analysis of the effects of permutation on the clustering.

\section{Latent Dirichlet Allocation}

The LDA model can be summarised as two matrices which establish the relationship between words, topics and documents. These matrices are constructed such that:

\begin{itemize}
	\item[] $\beta_k$ represents the word distribution for a given topic $k$
	\item[] $\theta_i$ represents the topic distribution for a given document $i$
\end{itemize}


This construction demonstrates how the process of LDA is an example of soft clustering, where a data point isn't strictly assigned to any one cluster. Under the LDA model, the words of the data set are given a probability of  being assigned to each possible topic - this explains the initial thought behind the construction of $\beta$ and $\theta$, to view how groups appear within individual documents. From this understanding, the matrix entities can be viewed as a decomposition of a matrix representing the word-document relationship for the entire corpus. Using this approach, an approximation for the 'ideal' total number of latent groups can be inferred by comparing their relative significance; see our implementation for how we chose to determine an appropriate number of topics.

As mentioned earlier, LDA is a 'generative' model, which corresponds to the assumption that an underlying probability distribution generates both the documents and latent topics. 

\noindent
Shown when constructing the matrices, LDA assumes that each row follows a symmetric Dirichlet distribution, with respective parameters $\alpha$, $\eta \in [0,\infty)$ expressed as:

\vspace{3pt}

\begin{center}
	$\theta_i \sim Dir(\alpha)$
	
	$\beta_k \sim Dir(\eta)$
\end{center}

For more information on the Dirichlet Distribution, see \citet{book}.
In the context of LDA application, it is unlikely that an individual document will consist of all the topics found in the corpus, so a parameter in the range $[0,1]$ is generally considered to be an ideal testing range. When fitting an LDA model to the provided dataset, the parameters $\alpha$, $\eta$ will need to be adjusted to reflect the word-topic and topic-document distributions found in the corpus data. Using the intuition behind their effect on the distributions produced, it is trivial to see that adjusting the $\alpha$ and $\eta$ parameters should determine the 'mixture' of topics in a document/words in a topic.

Figure \ref{fig:LDAPlateNotation} shows an overall view of the dependencies within the LDA model: 

\vspace{2pt}

\begin{center}
	\includegraphics[width=6.5cm]{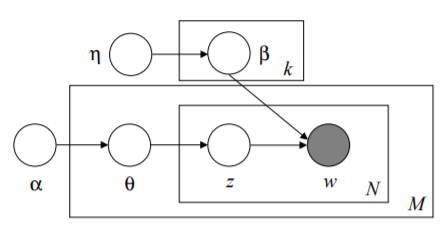}
	\captionof {figure} {Plate Notation for LDA  \cite{lda2003}}
	\label{fig:LDAPlateNotation}
	\begingroup
	\fontsize{8pt}{12pt}\selectfont
	
	\textit{Here, M denotes a document, N denotes the words contained within the document, k denotes the identified topics, and z denotes the topic(s) assigned to the observed word w. Each rectangle denotes a repetition within the model to form the overall structure of the corpus data.}
	
	\endgroup	
\end{center}

\vspace{7pt}

The main advantage when using LDA for topic modelling and clustering is the production latent groups which can be readily interpreted as topics. As a generative probabilistic model, it is possible to generalise an LDA model to classify documents outside of the training corpora. For a more in-depth look into the process of extending and training an LDA model into a classifier using the identified latent topics, see \citet{phan2008}, \citet{PAVLINEK201783}. 

The significant limitations of an LDA model arise from the inability to capture topic correlations (due to the independence assumed by using the Dirichlet distribution), the use with short-text documents, and the ambiguity around evaluation. For further reading, the problem surrounding correlations is directly approached in \citet{blei2007} where they propose the 'Correlated Topic Model'. The limitation encountered when using short-text documents stems from the lack of information available. By the nature of the data set being fitted to the LDA Model, as less information is provided, it is harder for the model to establish meaningful associations between documents, leading to documents within a topic generally being weaker in terms of 'similarity'. In terms of evaluating an LDA model, there are probability-based metrics available which provide a quantitative means by which the quality of any two topics models can be compared. However, as stated in \citet{HAGEN20181292}, 'the best-fitting model and human judgment are negatively correlated'. As a solution, an approach they detailed involved two steps. Firstly, a topic model is fitted by optimisation of some measure (e.g. perplexity) for each iterative step involved in the training method. Then humans would be required to manually assess the quality of the topics formed, allowing an optimal topic number to be derived that 'produces the best quality topics within the range [established from] the first step'. 

\subsubsection{Examples of Implementation}
The paper in which the LDA model was first proposed \cite{Pritchard945} used the model in the context of population genetics to gain insights into the population structure and assign individuals to groups according to indications from their genotypes. Due to the nature of the LDA model, it can be used in a copious range of applications, but naturally lends itself to the field of natural language processing. Some applications of LDA are:

\begin{itemize}
	\item Tag Recommendation System \cite{Kresteletala09}: Inspired by the tagging of online multimedia sources with irrelevant or niche tags; resources that have been tagged by users were used to find latent topics to map new resources to fewer, more relevant tags. In this case, each tag would be viewed as a 'word', and the tags associated to some resource as a 'document'.
	\item Review Analysis \cite{GUO2017467}: a data mining approach was used on a data set of 266,544 online hotel reviews for hotels across 16 countries. Using LDA, this approach identified different 'controllable dimensions' and their significance to inform hotels on how best to interact with their visitors.
	\item Fraud Detection \cite{XING20071727}: used in the telecommunications industry, the aim was to detect fraudulent calls from the high-volume network. LDA was used to develop profile signatures, so that any activity that deviated from what was considered 'normal' would be flagged for possible fraudulence.
\end{itemize}


\subsubsection{Determining an Optimum Topic Number}

Choosing the number of topics (K) was a significant challenge we faced as part of the implementation. For the LDA model, this parameter needs to be specified before the data set is fitted. Finding an optimal value tends to be found via experimental methods as a standard method to determine the optimal value doesn't exist \cite{Croft2009SearchE}.

When choosing the number of topics, it is important to consider the cases where K is less/more than the optimum number of topics. In the former case, the LDA model would produce over-generalised topics where data objects are over-aggregated, resulting in a model with clusters that don't produce any significant insights into a structure behind the data set. In the latter case, the LDA model would produce latent topics where very few data objects are clustered within each. This would reduce the possibility of the LDA model detecting an association between data objects and clustering them together. The extremes of both cases would irrespectively reduce the LDA model to an unusable state.

In rare cases, the context of application can dictate the optimum of cases. Since this wasn't the situation here, we decided to implement a Hierarchical Dirichlet Process (HDP) to provide an insight into the range of topic numbers we would take forward for further testing.
\section{Hierarchical Dirichlet Process}
A Hierarchical Dirichlet Process is a non-parametric Bayesian model for clustering grouped data. As suggested by \citet{HDP2006}, we view the HDP model as an extension of the LDA model for our implementation. When considering the construction, like we saw for LDA, the HDP model views the  documents of a corpus as a mixture of latent topics, and those latent topics as a mixture of words. The difference lies in how these models determine the number of topics.

Similarly to LDA, the HDP model assumes three matrices to establish relationships between words, topics and documents, constructed as below:

\begin{itemize}
	\item $\alpha_k$ represents top level Dirichlet variables sampled for a given topic $k$
	\item $\theta_j$ represents the topic distribution for a given document $j$
	\item $\varphi_k$ represents the word distribution for a given topic $k$
\end{itemize}

Thus the generative process for HDP is as follows, where these matrices are constructed with parameters $\beta, \gamma, \eta$:

$$\alpha_k \sim Dir(\gamma /K)$$
$$\theta_j \sim Dir(\eta \, \alpha_k)$$
$$\varphi_k \sim Dir(\beta)$$
$$z_{ij} \sim \theta_j, \, \, x_{ij} \sim \varphi_{z_{ik}}$$

\vspace{0.6em}

In this model, K denotes the number of topics, and for the purposes of HDP, is taken to tend to infinity. Like in the LDA model, each variable is modelled by a symmetric Dirichlet distribution, while the variable $z_{ij}$ is sampled from $\theta_j$ and each word $x_{ij}$ is sampled from the corresponding topic $\varphi_{z_{ik}}$. 

The LDA model requires a pre-determined parameter for the number of topics to cluster the data into, and models documents as a mixture of these fixed number of topics. On the other hand, the HDP model infers the number of topics by viewing this value as a variable generated by a Dirichlet Process. This process is 'hierarchical' in the sense that it adds another layer to the model, the Dirichlet Process, which determines the number of topics.

In the paper which first proposed the HDP model \cite{HDP2006}, an experiment is carried out to demonstrate the non-parametric nature of the HDP model, by evaluating these models when fit to a corpus of nematode biology abstracts. The two models were compared based on the evaluation metric 'perplexity'. For reference, the perplexity of a held-out abstract consisting of words $w_1,...,w_I$ is defined as
$$ \exp \bigg(-\frac{1}{I} \log p(w_1,...,w_I | \textrm{ training corpus} ) \bigg), $$
where $p(\cdot)$ denotes the probability mass function for a given model. The experimental findings of \citet{HDP2006} were that, on the basis of perplexity, the HDP model performed just as well as the best LDA model, and the posterior of the HDP model was consistent with that of the best LDA model.

This highlights the main advantage the HDP model has over LDA - the non-parametrised nature of HDP allows the consideration of the corpus data to consist of, potentially, an infinite number of topics, from which an optimum is inferred. Although this eliminates the challenge of determining an optimum topic number, the issue of required computational power arises to account for the possibly infinite feature of the model. Despite this, the non-parametric advantage is significant when dealing with data of an unknown nature, as it can be difficult to choose an optimal number of topics without full knowledge of the data.

\subsubsection{Examples of Implementation}
Its non-parametric nature allows the HDP model to be applied to various uses, some of which are:

\begin{itemize}
	\item Multi-population Haplotype Phasing \cite{Xing2006}: A new Bayesian approach to haplotype inference for multiple populations was developed, which incorporates HDP and has an improved accuracy to other haplotype inference algorithms.
	\item Word Segmentation \cite{Pei2013}: A refined model based on HDP was developed for word segmentation in the Chinese language, with improvements to the base measure using a dictionary-based model.
	\item Musical Similarity Computation \cite{Hoffman2008}: A method based on HDP for estimating the timbral similarity between recorded songs, which are represented as feature vectors.
\end{itemize}

\section{The Dataset}\label{sec:dataset}
Pre-processing of input data is a crucial part of any Machine Learning model. Our data set was obtained from \ifanon a course at our university\else{} the CM20254 Data Structures and Algorithms course offered at the University of Bath\fi. As part of the coursework, the students on this course were asked to submit three multiple choice question answer pairs related to various topics taught in the lectures. Furthermore, they were asked to review these questions in a peer assessed format. We focus our attention to only cluster the questions from this data set. The rationale behind this was to treat the data as if they were unanswered user queries, so that once clustered it would be easy to assign answers to specific clusters. This is part of a bigger project recently initiated at the University of Bath to help establish a clustering model that could be deployed on the online MSc courses. 

\subsection{Pre-processing of Data}

The data set contained over 1300 multiple choice question answer pairs (including their explanation), level of difficulty and overall quality score given by the students. As our data set of questions was in a .txt format, it was necessary to import it into a suitable format in order to be able to process the data, as well as retain its inherent structure. In order to do this, we used Python to clean the data set and convert it to .csv format which gave us a large degree of control over the structure of our final data set.

The next hurdle was to deal with questions that contained code scripts. This posed a problem with the available LDA models, as they the did not recognise code terms as potential tags for questions. We manually searched through the entire data set for code keywords and added respective tags at the start of those questions. The keywords tagged were: BigO, Modulo, for, if, while, else, print.

Prior to applying our model to the data set, we needed to make sure that it did not contain any grammatical errors. Since our data set contained various subject specific terms not found in the English dictionary, we manually searched for these terms and corrected any grammatical errors. 

The next step of preparing the data for the LDA model was removing punctuation and converting the data to lowercase text. This was easily implemented within the DataFrame structure using the 'sub' and 'lower' functions.

\section{Methodology}\label{sec:method}
To gain a broad approximation for the optimal topic number, the HDP model was deployed. When using the 'gensim' package, a truncation level had to be set, as it wouldn't be realistic to compute the Dirichlet Process over an infinite space. The upper bound for the parameter was set to be the size of the corpus data set; using this value would assume every question would occupy its own cluster. The lower bound for a topic within the distribution to be considered 'significant' was set to be the inverse of the corpus size - a considerably low boundary as this corresponds to the value taken by the uniform distribution for the topic range.

The topic parameter values found by this process were referred to as HDP-1. When investigating the number of topics within the distribution above the significance threshold, the topic numbers found were large enough that the LDA clusterings produced weren't able to form many significant associations between questions, leaving many questions unclustered. This suggested that the topic numbers of HDP-1 were over-approximations to the optimal value.

To narrow the scope of this approximation, we repeated the process above, but instead taking it a step further by increasing the significance boundary to the inverse of the HDP-1 value found. The number of topics within the distribution above the new boundary was referred to as HDP-2; the process was similarly extended to produce an HDP-3 value. Increasing the significance boundary certainly lead to a reduced value produced by 
HDP-2 and HDP-3. When used, the values produced for HDP-3 aggregated many questions into a few topics, and left a significant proportion of questions unclustered. This could be expected, as the boundary for HDP-3 became high enough that very few topics were above it - a strong sign that HDP-3 is an underestimate for the optimal value. By inspecting the clustering patterns from these HDP-produced values, it appeared as though HDP-2 was the most promising. Note that a larger data set could feasibly require further extensions of HDP-$x$ beyond what we found before a sensible optimal approximation can be found.

In an attempt to fine-tune our approximation for the optimal topic parameter value, an investigation was carried out into the number of topics that were being clustered within the LDA model. Using this reasoning, we formed a recursive process to improve the topic parameter. By using the number of topics that were used for clustering as the new topic value when fitting a new LDA model and recursively inspecting the effective number of topic clusters used, we observed the ratio of `topics used' to `topics specified' consistently converge to 1, over a multitude of tested runs. 

\subsubsection{HDP 1 vs HDP 2}

In our implementation, we make use of the Python package 'gensim' to form our HDP model. In the 'gensim' version of HDP, the model produces a probability distribution, detailing the likelihood of a given topic appearing in the ‘optimal’ set of topics. Initially, we used a minimum likelihood criterion of 1/n, where n was the size of our data set, as the minimum probability of a topic being significant. We then used this number of significant topics as a prediction to use within the LDA model. However, when running the LDA model with this HDP estimate for the number of topics, the clustering output was not optimised in all cases, as a number of topics were not being clustered to, for various values of n. Indeed, this problem seemed to worsen for larger data sets. For an initial corpus size of 100, the ratio of clustered topics to the initial number predicted by the HDP model was 1, which decreased to around 0.6 for a data set of more than 900 questions.

In order to overcome this issue, we explored using a stricter minimum significance criterion, and ran the HDP model a second time using 1/x, where x was the number of topics given in the first HDP estimate. This had the effect of yielding a smaller number of topics in all cases, as well as a more effective ratio, as seen in the second run in Fig. 1. The ratios in this case are all larger than 0.8, thus almost all topics are being clustered for various data set sizes, resulting in an improved clustering outcome from the LDA model.

\subsection{Zeno's Paradox}
When implementing the recursion process to fine-tune our topic parameter, an important consideration made was that of Zeno's Paradox \cite{sep-paradox-zeno}. Within our recursive method we were constantly checked for the efficiency of the LDA model i.e. the number of topics given as input compared to the number of topics actually used within the clustered output. Since these processes are probabilistic, there is always a slight chance that the efficiency ratio might decrease at points within the progression of the runs. To avoid encountering Zeno's paradox, we set up a subroutine, with initialised parameters $\gamma$ and $\eta$ such that the run would be ended if one of the following conditions were met:
\begin{itemize}
	\item The difference in efficiency ratio of successive runs is greater than $\gamma$.
	\item The number of successive runs with a steady decrease in efficiency ratio is greater than $\eta$.
	\item The ratio of effective topics hits 1.
	\item The efficiency ratio hits 1.
\end{itemize}
The first three points mentioned above help prevent entering an infinite loop (Zeno's Paradox). The final point is intended for exit from the recursive loop with a positive run. 

When carrying out the afore detailed iterative process over our question data set, we examined the impact of two specific variations: incrementally increasing the size of the questions data set being fitted with the LDA model and applying various permutations to the data set. In the first variation, we were expecting to observe some degree of trend when fitting the data set with the first 100, 200, 300,... questions. The second variation was implemented with the intention of observing the extent to which the questions involved in the various question sets would impact the LDA model that is fitted. Executing these variations required very different approaches. Testing and fitting the various question sets involved isolating the appropriate questions from the data set, which was readily achieved using available set functions commonly found in programming languages. Permuting the questions data set required making alterations to the raw data set before any pre-processing would take place.

Before any iterative runs could be carried out, the $\alpha$ and $\eta$ parameters had to be optimised for the specific data set that would be used. A range of possible values were obtained by maximisation of the coherence measure included in the 'gensim' package. This approach proved sufficient as these parameters were varied over the range [0,1] - a finite parameter space (justified earlier) as opposed to the infinite space for topics - allowing greater ease when identifying and inferring meaning from values of high coherence. The decision was made to use the default coherence measure, but further reading into the most effective topic measure implementations could start with \citet{10.1145/2684822.2685324}
\section{Empirical Results}
This section is predominantly split into two parts. In order to fully justify our method, we firstly need to validate that our methodology produces a better clustering for our data set. Further to this we need to show that our algorithm does provide a high efficiency in obtaining an optimal clustering. Our input data set was obtained from \ifanon our University\else the CM20254 course (Data structures and algorithms) taught at the University of Bath\fi. We then applied five different permutations on our data set to test the effects of it on clustering. On a broad spectrum, the clustering of the different permutations were similar.

The key differences to note were:
\begin{itemize}
	\item change in the ordering of significant topics 
	\item slight change in the number of clusters formed
	\item slight change in the keywords to a given topic.
\end{itemize}
We evaluated each topic within a cluster by qualitatively analysing the relevance of the questions present within the clustered topic. The relevance of a question was measured using the topic percentage contribution attribute, given by the LDA model. Topics with a large number of relevant questions were considered significant topics. 



The significant topics in HDP-2 are as follows:
\begin{itemize}
	\item Tree, Traversal, Binary
	\item Sort, Insertion, Stable
	\item Hashtable, Insert, Bucket
	\item List, Item, Number.
\end{itemize}

We have omitted listing all the keywords for each dominant topic, and selected those which best characterise the topic. The same dominant topics were obtained from our algorithm.
A noticeable difference was the increase of questions present in the significant topics. This was mainly due to a better clustering model.

\begin{exmp}
 The following questions were not part of the same cluster in the HDP-2 method.
 
 Q1278: Which of these sorting methods is slowest?
 
 Q465: Which sorting algorithm has the fastest average case?
 
 The options given with these questions have been omitted for pre-processing. Both these questions fall under the same category of the time complexity of sorting algorithms. Our algorithm detects this and clusters them under one topic.
\end{exmp} 

To illustrate the movement of questions throughout the process of our method, the following histograms detail the distribution of questions throughout the specified topics (the left column displays topics generated from HDP-2, while the right column displays the final topics reached with the iterative model).

\begin{figure}[htp]
	\centering
	
	\includegraphics[width=0.4\textwidth]{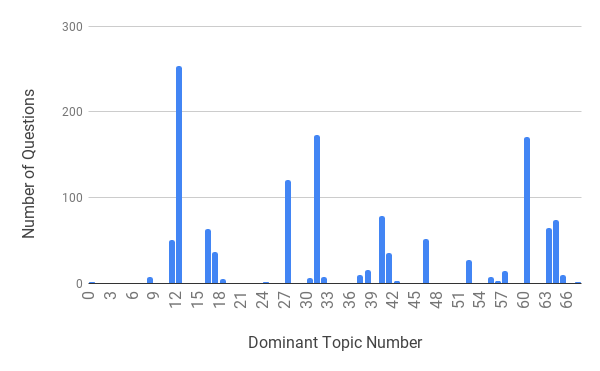}	

	\includegraphics[width=0.4\textwidth]{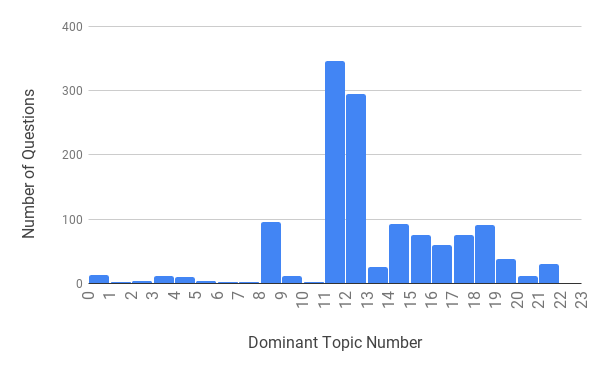}

	
	
	\caption{The pair of histograms correspond to the topic distribution of Permutation C using the LDA model with HDP-2 and the recursive model respectively.}
	\label{fig:Histograms}
\end{figure}

From the top histogram in Figure \ref{fig:Histograms}, observe topic 27 in  Permutation C, which contains the keywords 'complexity' and 'time'. The keyword ‘tree’, from topic 12 in permutation C (top histogram of Figure \ref{fig:Histograms}), remained as the core significant topic 12 in the final clustering (bottom histogram of Figure \ref{fig:Histograms}). In contrast, the questions contained in topic 27 with keyword ‘complexity’ were split over multiple topics, the most significant being topics 11 and 8, concerning ‘sort’ and ‘code’ respectively.
 
\subsection{Validation of Methodology and Results}
This subsection contains a detailed analysis for validating our methodology. We split each of our data sets i.e. the original data and its 5 permutations, into the first 100, 200,... questions. We then produced results from 100 runs on each of these data files using HDP-2 and our iterative model.

We started by recording the average estimated topics reached by our model against the HDP-2 model for each data set. The x-axis of the graphs below lists the number of questions contained in the data, while the y-axis contains the mean-mode metric. 

\begin{rmk}
	We computed the mean-mode by first computing the mode of the topic numbers. If multiple were found, we chose the closest value to the mean. Choosing the mean could result in obtaining a value that wasn't present within the set, resulting in a bad estimate of the output of our algorithm.
\end{rmk}

\vspace{-10pt}

\begin{center}
	\includegraphics[width=6.5cm]{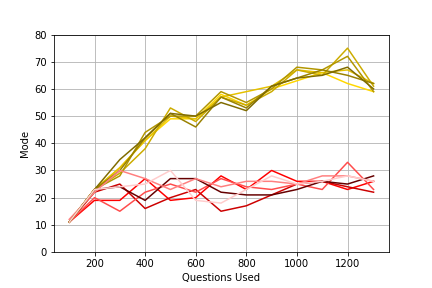}
	\captionof{figure}{}
	\label{fig:Mode Metric Comp}
\end{center}

The lines in yellow represent the topic numbers obtained from the HDP-2 model and the lines in red represent the final output of our model. The different shades in each colour represent the varied permutations of our data set. 

Notice that, for data sets containing a low number of questions (i.e. 100, 200, 300), the two models would roughly give the same output. This further justifies our previous claim that, for sufficiently small data sets, HDP-2 would give an adequate estimate to the optimal number of topics. 

\begin{figure}[htp]
	\centering
	
	\includegraphics[width=0.2318\textwidth]{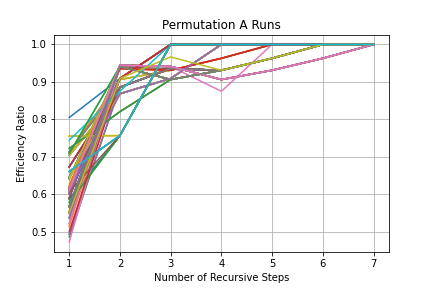}
	\hfill
	\includegraphics[width=0.2318\textwidth]{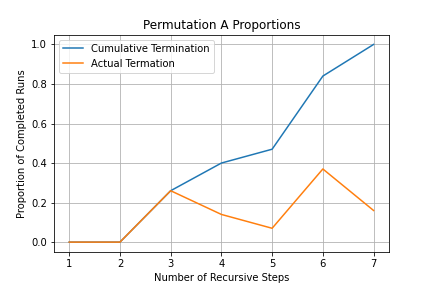}

	\medskip
	
	\includegraphics[width=0.234\textwidth]{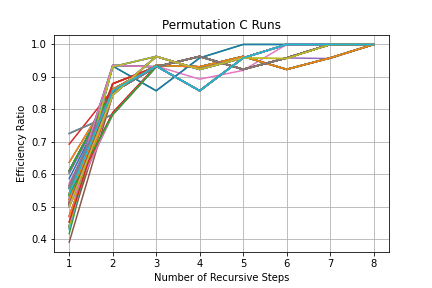}
	\hfill
	\includegraphics[width=0.234\textwidth]{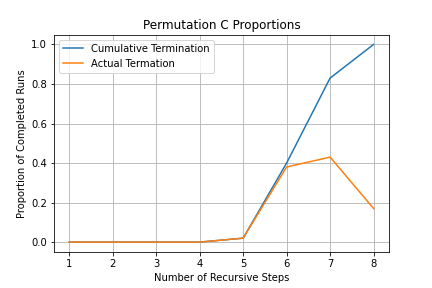}
	
	\medskip
	
	\includegraphics[width=0.234\textwidth]{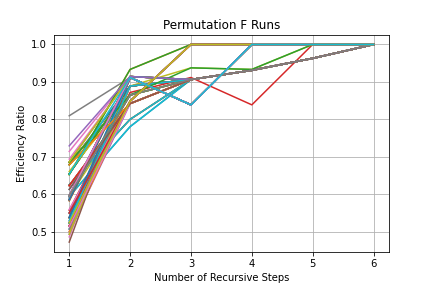}
	\hfill
	\includegraphics[width=0.234\textwidth]{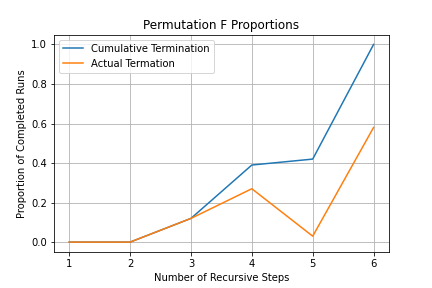}
	\caption{}
	\label{fig:Prop Ratio Collection}
\end{figure}

However, as the size of the data set increases, HDP-2 provides an excess number of topics. This can be further seen in the graphs in Figure \ref{fig:Prop Ratio Collection}.

\vspace{-5pt}

From the first column of Figure \ref{fig:Prop Ratio Collection}, which illustrates the evolution of the ratio for effective topics over 100 runs, we can see that the majority of runs saw a large improvement after the first few iterations, before reaching termination in the general range of 4 – 6 iterative steps, with Permutation C acting as the sole anomaly to this statement. Although we took into account the possibility of encountering Zeno’s Paradox, of the 100 runs for each of these permutations, there were no runs that ended in ‘Failure’. This result is promising, as it suggests that the chance of our algorithm outputting a failed run is extremely low. 

When observing the graphs in the second column of Figure \ref{fig:Prop Ratio Collection}, notice the distinction between the two lines. The blue line represents the cumulative proportion of the 100 runs that terminated at each step, while the orange line shows the marginal proportion that each step contributes. These graphs complement their corresponding run diagram well, as they identify any trends in path termination. Note that each graph finishes at 1, confirming that we have no failed cases. Similarly, the proportion graphs show that each set of runs terminates in the range of 4 – 6 iterative steps. These results show that within Permutation C, most runs terminated in 6 or 7 steps, confirming that our range is likely to be true.
Each range derived from the two figures provides an estimate of how many steps are required (on average) for our data sets. 
\begin{rmk}This approximation for the average number of iterative steps is dependent on the data set. This number would probably be different if another data set were to be used, but a similar process could readily be implemented to get a new range.\end{rmk}

Extending this further, we ran a time analysis on all six of our data sets, comparing the HDP-2 model vs our recursive model. 

\begin{center}
	\includegraphics[width=6.5cm]{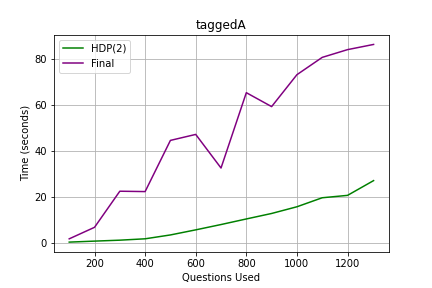}
	\captionof{figure}{}
	\label{fig:Time Results}
\end{center}

Figure \ref{fig:Time Results} clearly shows that the HDP-2 model has a slow exponential growth and our model displays a linear increase. This is an approximation, as our data set contains a maximum of 1300 questions. Given that the times are in seconds, the difference in timing of the two runs, although being 4-fold, is significantly small for practical use. Hence the trade off of a slower algorithm for a better clustering is justified. All the other data sets showed similar time graphs, further justifying that our model is not completely influenced by the ordering of questions-answer pairs.
\section{Conclusion and Further Work}
Exploring different models, the common approaches to clustering we came across were LDA, k-means and Support Vector Machines. Given that our data set was a combination of short pieces of text forming a large data set, traditional methods seemed to have a few shortcomings. For example, SVMs use binary classifiers, thus it would prove difficult to implement a series of SVMs to operate as a single system when classifying questions into a set of topics. It’s also worth mentioning that, depending on the size of the given data set, the scaling of a binary system for classification isn’t likely to be the most efficient. 
The k-means algorithm also presents certain drawbacks for clustering this type of data set. As it spatially clusters the data, it is heavily reliant on the Euclidean distance metric, as well as correctly selecting the number of cluster centers. This has the effect of incorrectly clustering data sets of varying cluster shapes.
A common drawback to the LDA model is associated with short texts; since short texts don’t provide as much information to the model, this makes the challenge of establishing associations and similarity between words and documents (via topics) significantly more difficult.

Despite these limitations, our model takes advantage of the clustering given by the LDA model and applies it recursively. The recursive nature of this algorithm helps attain the precision required to provide an effective clustering. Given that the LDA model is probabilistic, we test our methodology on a given data set over 100 runs. Our results showed a significant improvement as compared to the original model, specifically focusing on a lower number of cluster groups and more effective clusters. To verify that our model was independent of the ordering of the question-answer pairs in our data set, we applied our approach to 5 different permutations. All the results showed a significant similarity amongst the different permutations. The only key difference was obtained when we attempted to cluster the first few 100 questions for each data set. This was mainly due to the randomness of the data, which resulted in a different number of topics being present. This was normalised once the number of questions in the data set was over 500-600. Our method was slower than the original LDA model, however it is a drawback we can concede in order to produce a more efficient clustering.   

The data set we have used for our methodology has been trimmed down to only contain the questions from the question-answer pairs. Our next focus will be on understanding the influence of the answer part in the question-answer pairs of the data set. The main focus will be to provide a new method, based on our existing method, which could utilise the information given by the answer of a question in order to provide a more efficient clustering algorithm. One other interesting avenue we want to explore is the semantic meaning behind our clustering, inspired by the work done in \citet{Chang2009ReadingTL}. 
This will help us improve our existing model for a better clustering of question-answer pairs.

\bibliography{EAAI21_BIB}
\end{document}